\documentclass[sigconf]{acmart} 
\AtBeginDocument{%
  }

\setcopyright{acmlicensed}
\copyrightyear{2025}
\acmYear{2025}
\acmDOI{XXXXXXX.XXXXXXX}
\acmConference[CHI'25]{Workshop Purposeful XR: Affordances, Challenges, and Speculations for an Ethical Future }{April 26- 1 May, 2025}{Japan}
\acmISBN{978-1-4503-XXXX-X/2018/06}

\begin{document}


\title{Leveraging Metaphors in a VR Serious Game for Computational Thinking}

\author{I. Rodríguez}
%\email{webmaster@marysville-ohio.com}
\affiliation{%
  \institution{Departament de Matemàtiques i Informàtica. UBICS Research Institute. Universitat de Barcelona}
  \city{Barcelona}
  \country{Spain}
}
\email{inmarodriguez@ub.edu}

\author{A. Puig}
\affiliation{%
  \institution{Departament de Matemàtiques i Informàtica. IMUB Research Institute. Universitat de Barcelona}
  \city{Barcelona}
  \country{Spain}}
\email{annapuig@ub.edu}

\renewcommand{\shortauthors}{I. Rodriguez; A. Puig}

\begin{abstract}
  This paper presents Cooking Code, a VR-based serious game designed to introduce programming concepts to students (ages 12-16) through an immersive, scenario-driven experience. Set in a futuristic world where humans and machines coexist, players take on the role of a fast-food chef who must assemble food orders based on pseudocode instructions. By interpreting and executing these instructions correctly, players develop problem-solving skills, computational thinking, and a foundational understanding of programming logic. The game leverages the kitchen metaphor to teach computational thinking, using affordances for an immersive VR experience.
\end{abstract}

%\begin{CCSXML}
%<ccs2012>
%  <concept_id>00000000.0000000.0000000</concept_id>
%  <concept_desc>Do Not Use This Code, Generate the Correct Terms for Your Paper</concept_desc>
%  <concept_significance>500</concept_significance>
%  <concept_id>00000000.00000000.00000000</concept_id>
%  <concept_desc>Do Not Use This Code, Generate the Correct Terms for Your Paper</concept_desc>
%  <concept_significance>300</concept_significance>
%  <concept_id>00000000.00000000.00000000</concept_id>
 % <concept_desc>Do Not Use This Code, Generate the Correct Terms for Your Paper</concept_desc>
%  <concept_significance>100</concept_significance>
%  <concept_id>00000000.00000000.00000000</concept_id>
%  <concept_desc>Do Not Use This Code, Generate the Correct Terms for Your Paper</concept_desc>
%  <concept_significance>100</concept_significance>
%</ccs2012>
%\end{CCSXML}


\begin{teaserfigure}
  \includegraphics[width=\textwidth]{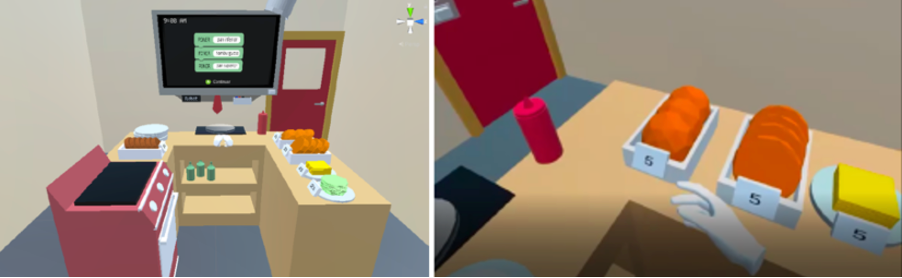}
  \Description{Screenshot of the Cooking Code VR game showing a player using virtual hands to interact with kitchen elements while solving programming challenges.}
  \caption{Cooking Code: VR serious game for computational thinking.}
  %\Description{Enjoying the baseball game from the third-base seats. Ichiro Suzuki preparing to bat.}
  \label{fig:VRGame}
\end{teaserfigure}

%\received{29 March 2025}

\maketitle

%1. Introducción
%    Presentar el desafío social elegido (enseñanza de programación a niños).
%    Explicar por qué es un problema relevante y qué barreras existen actualmente.
%    Introducir la metáfora de la cocina como enfoque educativo.
 %   Breve mención de cómo XR puede ayudar.

%2%. Problema y estado del arte
%Discutir la importancia de la enseñanza de programación en niños.
%    Explicar por qué los métodos actuales no son suficientes o tienen limitaciones.
%    Revisar investigaciones previas sobre enseñanza de programación con XR o metáforas educativas.

%3. XR como solución
%    Explicar qué affordances de XR (interacción espacial, inmersión, etc.) se aprovecharán.
%    Describir cómo la metáfora de la cocina y XR pueden mejorar la enseñanza de programación.
%    Ejemplos concretos de interacción en el juego.

%4. Desafíos tecnológicos y futuros avances
%    ¿Qué avances en XR se necesitan para mejorar esta solución?
%    ¿Qué barreras existen en hardware, software o diseño educativo?
%5. Consideraciones éticas y de implementación
%    Posibles limitaciones y problemas (ej. accesibilidad, costos).
%    ¿Cómo aseguramos que sea una herramienta inclusiva y efectiva?
%6. Conclusión
%    Resumen de cómo XR puede ayudar a este problema social.
%    Propuesta de próximos pasos en investigación/desarrollo.

\section{Introduction}
Serious games, which blend entertainment with educational objectives, have demonstrated their potential to enhance engagement and knowledge acquisition \cite{gee2003video} \cite{wouters2013meta}. Among them, serious games designed to foster computational thinking (CT) have gained attention for their ability to teach fundamental programming concepts through interactive and problem-based learning \cite{su2023systematic}.

Despite the increasing importance of CT skills in today’s digital world, traditional programming education remains challenging for many students. Apart from visual approaches such as Scratch \cite{scratch}, many others often rely on abstract explanations and text-based programming environments, which can be intimidating and discouraging for beginners. 

Studies have shown that game-based learning can help mitigate these difficulties by providing engaging, context-driven learning experiences \cite{relkin2021learning}. Moreover, virtual reality (VR)—with its immersive and spatial interaction capabilities—offers an opportunity to further enhance serious games by increasing engagement, presence, and hands-on learning experiences. Indeed, previous studies on VR for learning in general \cite{lampropoulos2024virtual} \cite{villena2022effects}, and programming education in particular \cite{gerini2023gamified} \cite{bueno2022effects}, have shown promising results in improving students' motivation and comprehension.

This paper presents Cooking Code\footnote{The game was developed using the Unity game engine, using the XR Interaction Software Toolkit and Meta Quest 2 as the head-mounted display}, a VR-based serious game designed to introduce students (ages 12–16) to programming concepts through an immersive, scenario-driven experience. We contribute with a novel design for teaching computational thinking by using VR affordances (grabbing, moving, sensorial feedback) within a kitchen metaphor. The game is set in a futuristic world where humans and machines coexist. Players take on the role of a fast-food chef who must assemble food orders written on pseudocode instructions provided by the machines. By interpreting and executing these instructions correctly, players develop problem-solving skills, computational thinking, and a foundational understanding of programming logic.

%Cooking Code teaches computational thinking by using VR affordances within a kitchen metaphor. Real-world cooking processes are mapped to programming concepts (like ingredients as variables and programs as recipes), and interactive elements like utensils and ingredients intuitively guide players in problem-solving and logical structuring.

\section{Related work}

Several research studies as well as commercial platforms have explored computational thinking in VR. Pierre et al. \cite{pierre2020learning} conducted a study examining student acceptance and the efficacy of VR games in programming education, revealing positive results attributed to the simulation of real-world scenarios and interactive elements. 

CoSpaces \cite{delightex} \cite{khattib2024quasi} is an online platform that enables users to create, explore, and share 3D worlds, with support for virtual reality interaction. Similar to the well-known Scratch platform \cite{scratch}, CoSpaces is designed for beginners, allowing users to build virtual experiences without extensive programming knowledge. Projects can be developed using Blockly (a visual programming language) or JavaScript, making it a flexible tool for educational purposes \cite{weintrop2019block}. Unlike many VR educational applications that focus primarily on passive visualization (e.g., virtual museum tours), CoSpaces emphasizes interactive content creation. 
%Several other studies have explored how VR can enhance computational thinking and programming education. 
%GamiCAD (Xu et al., 2020) is a VR-based system that teaches programming through a gesture-based coding environment, allowing users to manipulate virtual objects to construct algorithms. 

VRCoding, a recent work based also on VR block coding \cite{gerini2023gamified}\cite{gerini2023d}, aims to teach computational thinking immersively by utilizing passive haptics (i.e., instead of relying on active haptic devices that generate forces or vibrations, passive haptics uses the inherent physical properties of objects) for enhanced interaction. User testing with secondary school students demonstrated positive feedback regarding presence and experience compared to standard monitor-based coding. 
%These approaches inspired aspects of our game, Cooking Code, particularly in incorporating visual programming within an immersive environment.

In the university context, a VR game, Imikode \cite{sunday2023usability}, was designed to teach object-oriented programming by building a 3D virtual world of  objects such as trees, houses, and animals through a storytelling approach. The findings suggested that VR's immersion and direct visualization of abstract concepts motivate students and significantly enhance their programming comprehension and skills. 

Similarly, Tanielu et al. developed OOPVR \cite{tanielu2019combining}, a VR experience using analogies like houses and blueprints, to clarify abstract OOP concepts for students. Pre- and post-surveys showed a significant increase in students' confidence in visualizing OOP concepts after using OOPVR.
Our research uses block-based characteristics and analogies from previous studies, but differentiates itself by implementing a novel and
intuitive metaphor that enables a seamless introduction to programming.

Nevertheless, despite the benefits, the recent literature reports that VR serious games raise ethical concerns regarding user data, immersion, accessibility, and safety. Both HCI \cite{skulmowski2023ethical} and AI \cite{zhong2025computational}  communities are addressing these concerns including, for example, data security \cite{iqbal2023security}, respectful interactions of AI companions \cite{rosello2022ethical} and physical risks like motion sickness \cite{wang2024effects}.

%Similarly, XRBlocks, a visual programming language for XR-based coding , aims to enhance computational thinking education through immersive VR environments. Employing passive haptics and real-time tracking, the system provides tactile feedback and interactive learning experiences. Designed for introductory programming, XRBlocks has demonstrated potential in increasing student engagement, particularly within STEM subjects, through interactive arcade game-based exercises.

%Cooking Code builds upon previous work by integrating structured pseudocode interpretation with VR-based interactivity, offering an approach that combines hands-on engagement with programming logic comprehension.

%del tfg de david
%Owlchemy Labs' Job Simulator, a VR game simulating humorous real-world professions, offers a playful and immersive experience through realistic VR interactions \cite{job_simulator}. While not explicitly educational, its game design, interaction mechanics, and environmental modeling served as a key reference for Cooking Code. Cooking Code draws inspiration from Job Simulator's engaging VR controls, adapting them to create a structured learning experience focused on computational thinking.

\section{The kitchen metaphor and related affordances in VR}

Cooking Code game leverages the kitchen metaphor to teach computational thinking, using affordances for an immersive VR experience. The kitchen metaphor aligns real-world cooking processes with programming concepts (ingredients as variables and programs as recipes), while affordances in VR, such as interactive utensils (grill, plate, tray) and ingredients, intuitively guide players through challenge-solving.

\subsection{Game overview}
\begin{figure}
    \centering
    \includegraphics[width=0.8\linewidth]{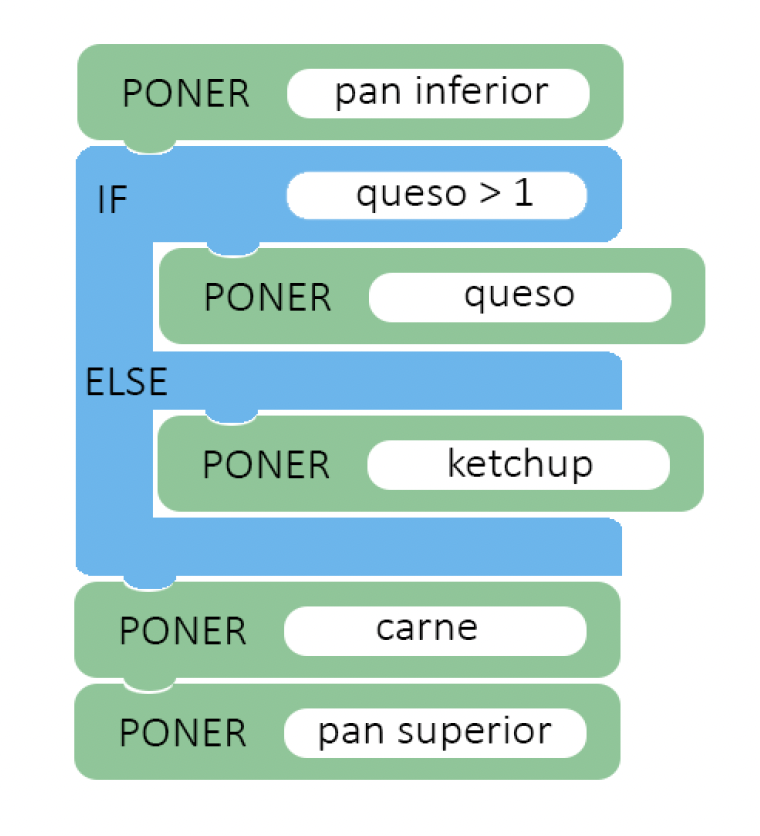}
    \caption{Food order: a hamburger with conditional logic. Spanish-English translation: PONER-PUT, pan inferior-bottom bread, queso-cheese, carne-meat, pan superior-top bread}
    \Description{Food order: a hamburger with conditional logic. Spanish-English translation: PONER-PUT, pan inferior-bottom bread, queso-cheese, carne-meat, pan superior-top bread}
    \label{fig:order-conditional}
\end{figure}

The design goal is to evoke fun and curiosity about programming through interactive, hands-on gameplay. 
%The game simplifies programming concepts in a fun, engaging way, allowing players to solve problems and gain a sense of achievement.
Set in a futuristic world where languages and programming syntax have merged, the game immerses players in a unique narrative. The player takes on the role of a cook in a burger restaurant, tasked with fulfilling incoming orders. Orders are displayed as pseudocode (see Figure \ref{fig:order-conditional}), requiring players to interpret and assemble the burger accordingly. The design of the orders is based on the visual programming found in platforms like Scratch or CoSpaces. Each of the actions to be performed, or instructions, is represented by a block. The order, in the end, is a set of blocks that fit together in a specific sequence, i.e. the recipe followed by the cooker (player) to satisfy the order.

\begin{figure}
    \centering
    \includegraphics[width=0.8\linewidth]{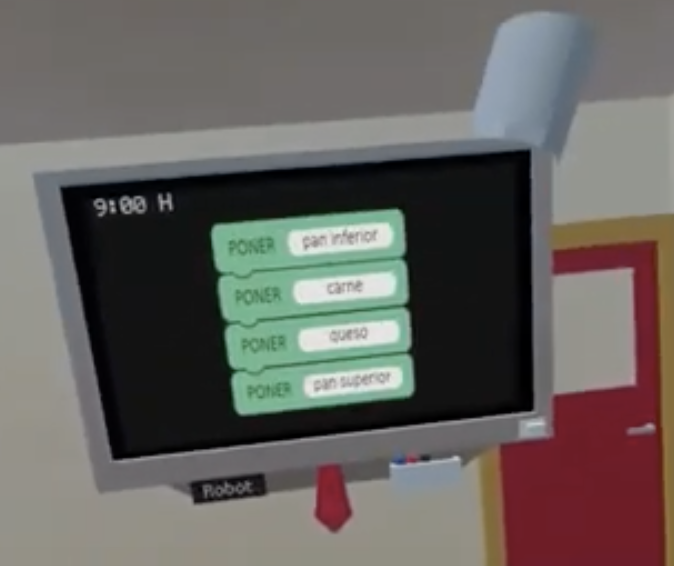}
    \caption{TV where feedback is shown, adorned with a cooker's hat and a tie}
    \Description{TV where feedback is shown, adorned with a cooker's hat and a tie}
    \label{fig:tv}
\end{figure}

Ingredients (lettuce, ketchup, cheese, top and bottom bread) are limited, requiring the player to use conditional logic. Each ingredient has a defined quantity that resets at the start of a new workday, which also displays earned experience in a TV screen featuring a cooker hat and tie (see Figure \ref{fig:tv}). The player must complete as many burgers as possible within the day, tracked by an in-game clock. Some ingredients need specific actions, such as cooking the meat before use. Taking advantage of immersive audio, the player hears the cooking sound and sees smoke as the meat cooks.
%Since ingredients are not infinite, players must restock from their respective containers, reinforcing resource management.

While solving the challenge, the player's core interactions are grabbing and placing — pick up ingredients and assemble them by placing them in the virtual plate. Once the burger is complete, the player delivers it to the serving tray.  The game’s VR affordances make these interactions intuitive, simulating real-world cooking actions while teaching computational thinking.

Each time players complete an order, they receive immediate feedback through a spoken message, also displayed in the TV. These messages are designed to be clear, concise, and unambiguous to facilitate learning. Given the target audience, the tone is kept simple and encouraging, avoiding overly formal language. Effective feedback not only helps players recognize mistakes and prevent frustration but also enhances motivation and engagement. 
%In an educational game, clarity is crucial to reinforcing learning and maintaining a positive experience.

\begin{figure}
    \centering
    \includegraphics[width=0.8\linewidth]{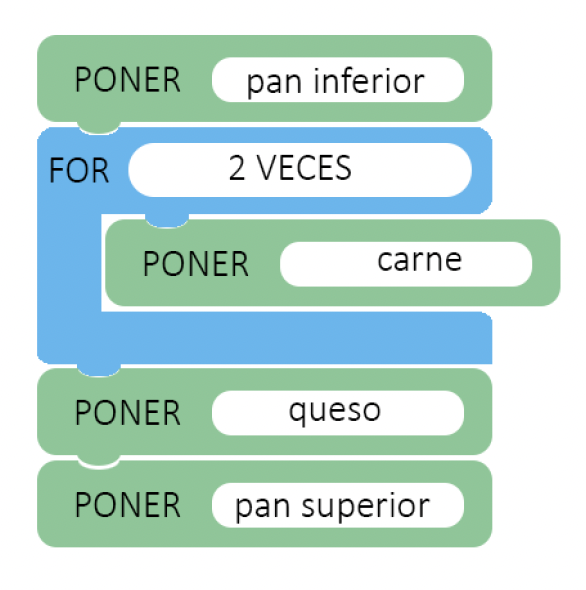}
    \caption{Food order of a double hamburguer. Iterative logic. Spanish-English translation: PONER-PUT, pan inferior-botton bread, carne-meat, 2 VECES-2 REPETITIONS, pan superior-top bread}
    \Description{An image with food order of a double hamburger}
    \label{fig:order-iterative}
\end{figure}

\begin{figure}
    \centering
    \includegraphics[width=0.4\linewidth]{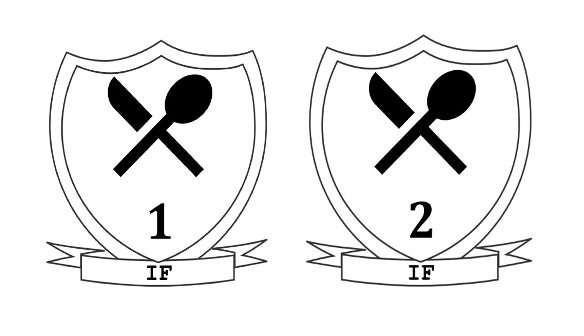}
    \caption{Achievements for completing 1 and 10 conditional (IF) food orders}
    \Description{An image with achievements for completing 1 and 10 conditional (IF) food orders}
    \label{fig:achievement}
\end{figure}

Over multiple sessions, players  aim to achieve high scores and accumulate skills. The ultimate goal is to complete all challenges and earn every achievement. Progression is driven by skill improvement and unlocked achievements (see Figure \ref{fig:achievement}). Complexity increases as new ingredients and recipes are introduced incorporating sequential, conditional and iterative logic (see Figures \ref{fig:order-conditional} and  \ref{fig:order-iterative}, respectively), reinforcing learning through gradual challenges. Note that the design of the achievement incorporates kitchen utensils as a visual metaphor to communicate a cooking-related accomplishment.

%\begin{figure}
%    \centering
%    \includegraphics[width=0.8\linewidth]{images/order.png}
%    \caption{Order of an hamburguer: sequential logic. Spanish-English translation: PONER-PUT, pan inferior-botton bread, hamburguesa-hamburguer, pan superior-top bread}
%    \label{fig:order}
%\end{figure}

%The game presents obstacles that adapt to the player's skill level, ensuring difficulty gradually increases through more complex orders and additional ingredients. Periods of calm are incorporated to highlight progress. Rewards include instant feedback such as customer reactions, as well as badges and medals for successful order completion, allowing players to track their achievements.

\subsection{Design of the playing (cooking) area}

\begin{figure}
    \centering
\includegraphics[width=1.0\linewidth]{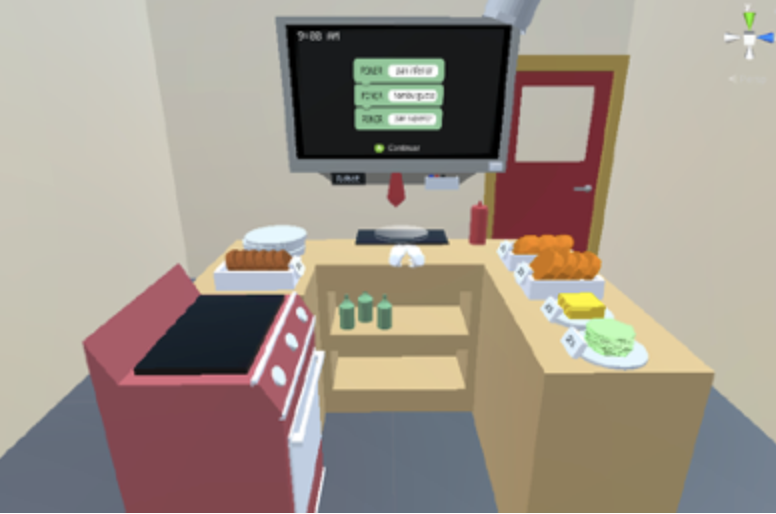}
    \caption{Design of the playing zone: cooker's table}
    \label{fig:table}
    \Description{An image with a cookers table with utensils}
\end{figure}

The cooking area was designed to ensure smooth and intuitive gameplay, minimizing unnecessary or abrupt player movements. Elements are grouped based on their usage order and position. On one side of the table (see Figure \ref{fig:table}), all ingredients except the meat are arranged, with the most frequently used ones  — top and bottom bread — placed closest to the player to reduce movement. The meat is positioned next to the grill for a natural and efficient cooking flow. In the center, plates and ketchup are placed for easy access at the beginning or end of an order.

%\begin{figure*}
%    \centering
%    \includegraphics[width=0.9\textwidth]{images/IngredientGenerationSelection.png}
%    \caption{Containers with ingredients. From left to right: ingredients, selected ingredient and grabbed ingredients }
%    \label{fig:selection}
%\end{figure*}

\begin{figure}
    \centering
    \includegraphics[width=1.0\linewidth]{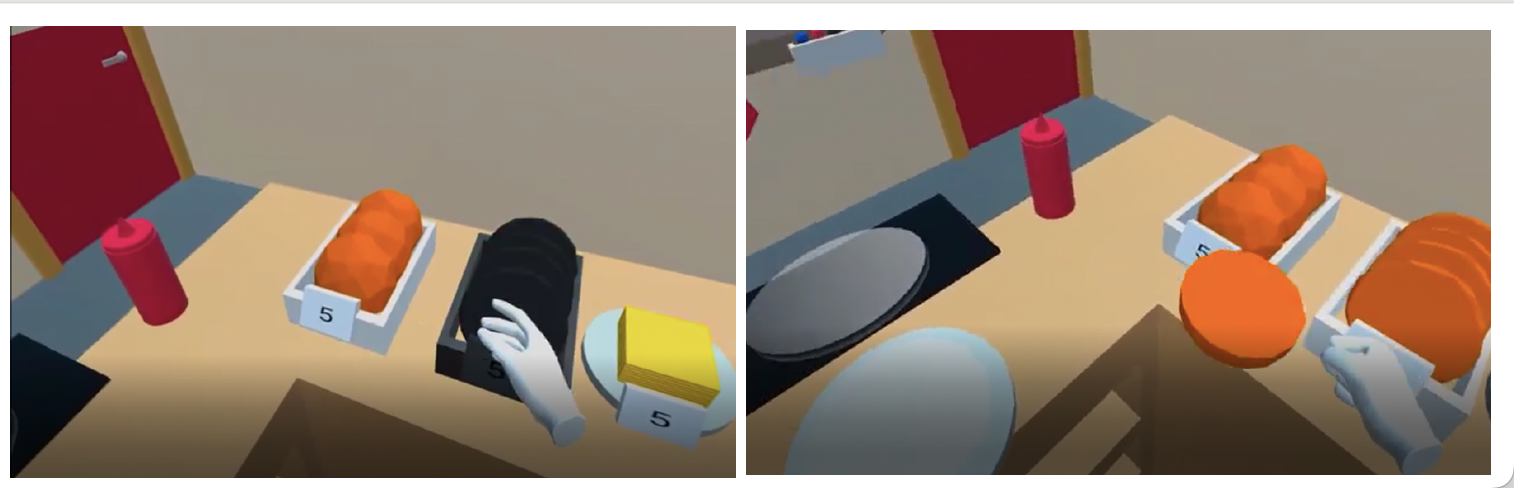}
    \caption{Containers with the ingredients. Left: selected ingredient, Right: grabbed ingredient}
    \Description{An image with the containers of the ingredients}
    \label{fig:selection}
\end{figure}

The delivery area (tray where the cooked hamburger is placed) went through several design iterations. Initially placed on the side to keep the workspace clean, it required too much head-turning, disrupting order tracking. A raised placement was also tested but felt unnatural. Ultimately, positioning it directly in front of the assembly area allowed for continuous order visibility, intuitive VR movement, and immediate feedback checking.

Regarding the grabbing interaction for ingredients selection. Figure \ref{fig:selection} shows the ingredients' containers. They afford easy selection, with numbers clearly indicating the quantity of each ingredient. As the user's hand approaches a black overlay highlights the selected tray, (see image in the middle of the Figure) providing visual feedback. Finally, the third image shows the player successfully grabbing an ingredient, further demonstrating the intuitive affordance of the virtual hand and its ability to manipulate objects within the VR environment. This interaction relates to embodiment \cite{pyasik2020visual}, contributing to the user's sense of agency and body ownership.

%resumen del juego
%The game begins with the presentation of a specific order that the player must fulfill by selecting and combining ingredients on a plate. The game interface is inspired by visual programming, where actions are represented by blocks that the player must arrange in the correct sequence to complete the order. Each selected ingredient is checked for availability, and only available ingredients can be added to the plate. Once the player believes the hamburger is complete, they deliver it. The game checks if the delivered hamburger matches the original order. If the hamburger is correct, the player receives positive feedback and experience points. If it's incorrect, they receive negative feedback and no points are awarded. The game also incorporates a time limit, adding an element of challenge. After each order, the game provides feedback and presents a new order, continuing the loop until time runs out. The game design emphasizes clear feedback to facilitate learning, using simple language and encouraging messages to maintain player engagement.

\section{Conclusions}

Through Cooking Code VR game, this research investigates how a metaphor along with VR affordances such as immersion, embodiment, and spatial interaction can address the limitations of window-text-based programming education. Players assume the role of a burger restaurant cook, fulfilling food orders presented as pseudocode, which they must interpret to assemble the burgers. Our on-going work is focused on an evaluation of the game with students, and a multi-user study of space configuration in competitive and cooperative scenarios. Other technological advancements such as multisensory experiences (e.g., scent dispensers in VR environments), generative artificial intelligence to dynamically create challenges, and personalization and customization are interesting research directions. 

%ethical considerations associated with immersive educational environments.

\begin{acks}
This work was supported by the SGR CLiC project (2021 SGR 00313), FairTransNLP-Language (PID2021-124361OB-C33, and ACISUD (PID2022-136787NB-I00 funded by MICIU/AEI/10.13039/501100011033).
\end{acks}

\bibliographystyle{ACM-Reference-Format}
\bibliography{sample-base}


\end{document}